\documentclass[a4paper, 10pt, conference, one column]{IEEEtran}

\usepackage[latin1]{inputenc}
\usepackage[T1]{fontenc}
\usepackage[english]{babel}

\usepackage[pdftex]{graphicx}
\DeclareGraphicsExtensions{.pdf,.jpeg,.png}
\graphicspath{{figures/}{}}

\usepackage[cmex10]{amsmath}
\usepackage{amssymb}
\usepackage{trfsigns}
\usepackage{eurosym}
\usepackage{empheq}%
\usepackage{cite}
\usepackage[nolist]{acronym}

\usepackage{thmtools}
\usepackage[detect-all]{siunitx}
\usepackage{booktabs}
\usepackage{multirow}
\usepackage{threeparttable}

\usepackage[version=3]{mhchem}  

\newcommand{\spacepipe}{\,|\,}  

\newcommand{\SoC}[1][k]{x^{\textnormal{SoC}}_{#1}}

\DeclareSIUnit{\EUR}{\text{\euro}}

\usepackage{xfrac}

\title{PV Integration in Low-Voltage Feeders\\ with Demand Response}

\author{
  \IEEEauthorblockN{Xiangkun Li, Theodor Borsche and G\"oran Andersson}
  \IEEEauthorblockA{Power Systems Laboratory, ETH Z\"urich\\
    lixi@student.ethz.ch, borsche\spacepipe andersson\,@\,eeh.ee.ethz.ch
  }
}

\begin{document}
\begin{acronym}[swissgrid]
\acro{OLTC}{On-Load Tap-Changing Transformer}
\acro{AMI}{Advanced Metering Infrastructure}
\acro{SM}{Smart Meter}
%
\acro{RES}{Renewable Energy Sources}
\acro{PV}{Photo-Voltaic}
%
%
\acro{DSO}{Distribution System Operator}
\acro{TSO}{Transmission System Operator}
%
%
\acro{LV}{Low-Voltage}
\acro{MV}{Medium-Voltage}
\acro{HV}{High-Voltage}
\acro{HVDC}{High-Voltage Direct Current}
%
\acro{BG}{Balance Group}
\acro{BE}{Balancing Energy}
%
\acro{ACE}{Area Control Error}
\acro{AGC}{Automatic Generation Control}
%
\acro{DR}{Demand Response}
\acro{DSM}{Demand Side Management}
\acro{EWH}{Electric Water Heater}
\acro{HVAC}{Heating, Venting and Air-Conditioning}
\acro{PLC}{Power Line Communication}
%
\acro{SOC}[SoC]{State of Charge}
\acro{SOH}[SoH]{State of Health}
%
\acro{MPC}{Model Predictive Control}
\acro{RHC}{Receding Horizon Control}
\acro{LP}{Linear Program}
\acro{QP}{Quadratic Program}
\acro{MILP}{Mixed-Integer Linear Program}
\acro{sLP}[SLP]{Stochastic Linear Program}
%
\acro{PF}{Power Flow}
\acro{OPF}{Optimal Power Flow}
%
\acro{EKZ}{the utility of the Kanton Zurich}
\acro{CAISO}{California Independent System Operator}
\acro{ERCOT}{Electric Reliability Council of Texas}
\acro{swissgrid}{Swiss Transmission System Operator}
\end{acronym}
\maketitle

\begin{abstract}
Increased distributed \ac{PV} generation leads to an increase in voltages and unwarranted backflows into the grid. This paper investigates \ac{DR} with \acp{EWH} as a way to increase the \ac{PV} hosting capacity of a low-voltage feeder. A control strategy relying only on power measurements at the transformer is proposed. Flexible loads are optimally dispatched considering energy acquisition costs, a \ac{PV} shedding penalty, and power and energy constraints. Furthermore, grouping of loads and \ac{PV} plants is investigated, and switching penalties are used to reduce the unnecessary switching of loads. It is shown that this strategy can substantially increase the \ac{PV} hosting capacity of a \ac{LV} feeder, even when only basic controllability is available.
\end{abstract}

\acresetall

\section{Introduction}

%

With over \SI{130}{\giga\watt} installed worldwide~\cite{EPIA2013}, \acp{PV} are currently the third most dominant form of renewable energy, behind hydropower and wind, in terms of capacity. The average annual growth rate of PV has been \SI{60}{\percent} over the last five years~\cite{REN21_2013}. Due to strong policy support, \ac{PV} was projected to surpass wind as the fastest growing source of electricity generation for the first time in 2013~\cite{Bloomberg2013}.

However, grid integration of massive amounts of distributed, stochastic generation has introduced many technical challenges for power system operators. In the case of \ac{PV}, one key issue that arose in Europe is voltage constraint violations in distribution networks. Approximately \SI{80}{\percent} of \ac{PV} power is fed into the \SI{400}{\volt} \ac{LV} grid~\cite{Ernst2012}. When generation in-feed exceeds local demand, voltage levels rise, causing damage to consumer and grid equipment if no action is taken. Safety concerns are also of importance as protection equipment now need to take into account bidirectional power flows.

Staying within the allowed voltage band is a greater challenge in \ac{LV} systems due to a comparatively smaller short circuit impedance and a larger \sfrac{R}{X} ratio. This often causes voltage violations before line limits are reached~\cite{Degner2011}. 
With ambitious renewable energy targets to meet in the next few decades, many solutions have been proposed to increase the \ac{PV} hosting capacity of \ac{LV} grids.
Seven key ways are outlined below. The first two methods aim to increase the system's ability to transfer power to higher voltage levels, while the remaining methods attempt to handle the surplus power locally~\cite{Bucher2013PVSEC}:
\begin{LaTeXdescription}
  \item[Reactive power control]
    Consuming reactive power while injecting active power limits voltage rise, but increased reactive power flows may create voltage issues in the medium voltage network and lead to higher losses.
  \item[\ac{OLTC}] 
    Lowering the secondary transformer voltage may not be possible if many lines are connected to a single bus and require a constant voltage.
  \item[Correlation with load profiles] 
    While assuming zero correlation is too conservative, central Europe exhibits poor load correlations due to low levels of air conditioning.
  \item[Active power curtailment]
    Key method implemented today due to its simplicity and allowance for emergency control.
  \item[Orientation change of \ac{PV} installations]	  
    A smoother generation curve is obtained when peak production periods are shifted, but this method currently lacks economic incentives as producers are only paid for energy generated.
  \item[Decentralized storage]	  
    Incentives for self-supply and private storage exist, but implementing peak production storage depends on future regulatory and costs structures.
  \item[\ac{DR}] 	  
    Power and energy curtailment are minimized and local energy use is prioritized, potentially reducing losses and system congestion. 
\end{LaTeXdescription}
Grid reinforcement and expansion is also always an option, but should be avoided if possible due to high costs and a lack of exploitation of existing resources.

This paper investigates the potential of \ac{DR} to increase the \ac{PV} hosting capacity of a \ac{LV} network. The charging schedule of \acp{EWH} is optimized to manage \ac{PV} generation peaks in order to limit the maximum voltage rise of the line to \SI{3}{\percent} as per DACHCZ code specifications~\cite{DACHCZ2007}. The control strategy relies only on power measurements at the transformer.
Uncertainties associated with weather forecasts are taken into account using model predictive control. Cost structures are investigated, as well as limited controllability due to the grouping of assets and the trade-off between switching frequency and PV in-feed shedding. Simulations are run on real data from Swiss households and \ac{PV} installations.

The paper is organized as follows: Section~\ref{sec:description} gives an overview of the test system and the data used. Section~\ref{sec:control} describes the control scheme for improving the \ac{PV} hosting capacity with \ac{DR}. Results are presented in Section~\ref{sec:results} and discussed in Section~\ref{sec:discussion}, before Section~\ref{sec:conclusion} concludes.

\section{System Description}\label{sec:description}

%
%
%
%
\subsection{Grid Topology}

\begin{figure}[ht!]
	\centering
	\includegraphics[width=80mm]{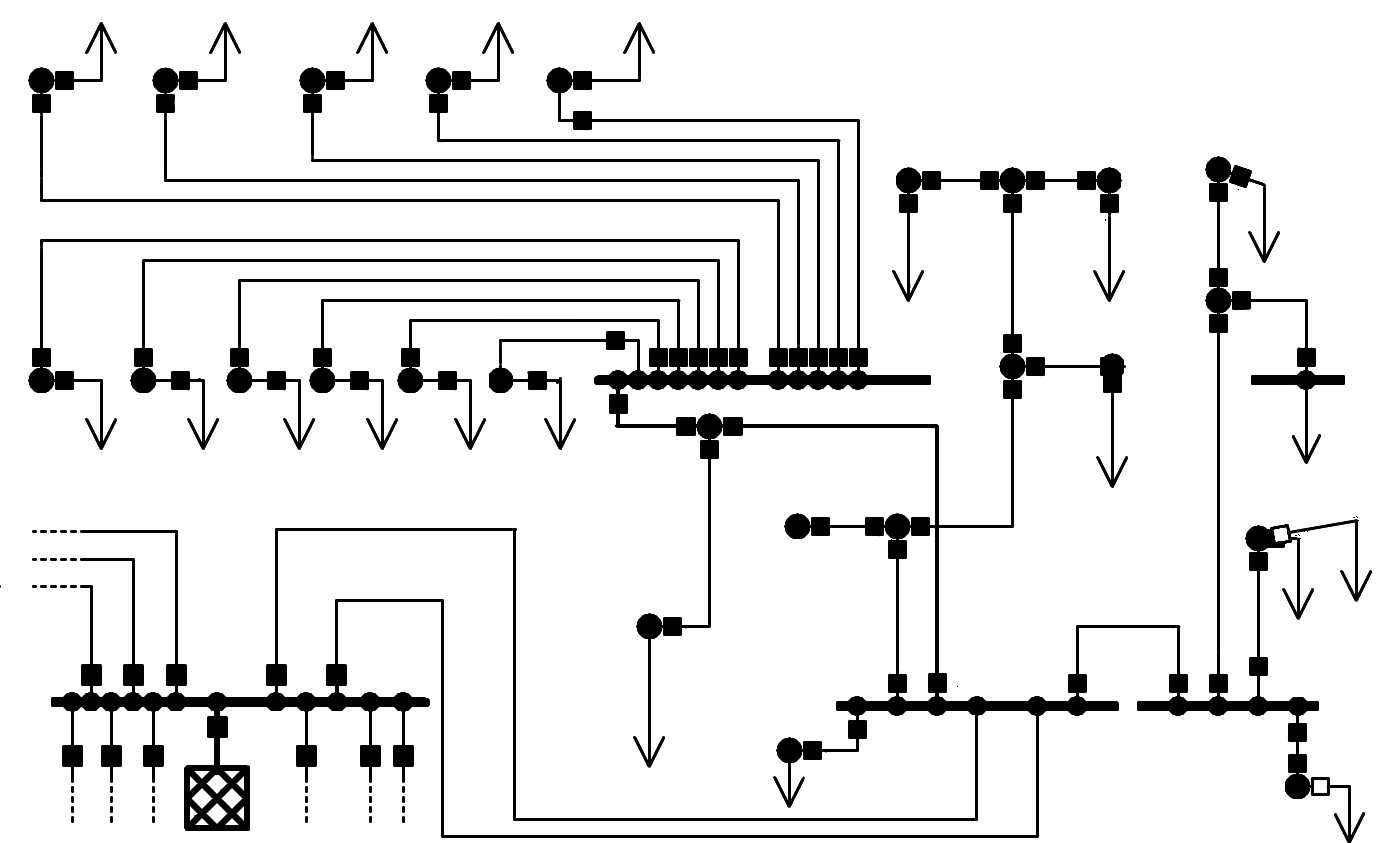}
	\caption{Grid topology of the \SI{400}{V} test system}
	\label{fig:grid_topology}
\end{figure}

The grid topology used is based on a representative Swiss suburban \SI{400}{\volt} distribution network (Figure~\ref{fig:grid_topology}). Two \SI{630}{\kilo\volt\ampere} transformers connect the \SI{400}{\volt} \ac{LV} grid to the \SI{16}{\kilo\volt} \ac{MV} grid. Two of the eleven feeders connected are considered here. This double-feeder supplies a residential area with twenty nodes. PV installations are assumed to be evenly distributed at the load buses. Voltage at the main busbar is fixed to 1.01\,p.u. and cannot be changed as it might lead to voltage limit violations in the remaining nine feeders.
 
\subsection{Measured Data}
\ac{PV} production profiles for active and reactive power are obtained from two installations rated at \SI{61}{\kilo\watt_p} and \SI{30}{\kilo\watt_p}, with an average capacity factor of \SI{9.83}{\percent}. Load information is extrapolated from measurements taken from 53 Swiss households. Connections are determined based on average transformer loading considerations and the maximum allowable voltage drop during peak periods. The simulation incorporates 200 households, with 10 households per bus, where minimum voltage levels never drop below~0.96\,p.u.
Measured data for load and \ac{PV} production is available from January 1st, 2013 to December 31st, 2013 with a temporal resolution in minutes. Missing data is linearly interpolated when missing values span less than 15 minutes. Days with more than 15 consecutive minutes of missing data are eliminated from the simulation.

\subsection{Hot Water Consumption Profile}
\begin{figure}[htb]
	\centering
	\includegraphics{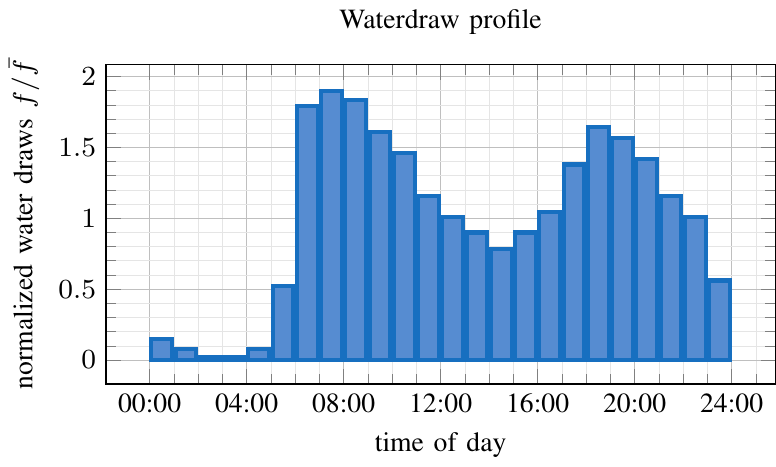}
	\caption{Daily hot water consumption profile}
	\label{fig:hot_water}
\end{figure}

The state of charge of the \acp{EWH} are assumed to be known at all times. An initial state is chosen randomly, and later states are computed using a simple hot water demand model that maps Swiss consumption rates onto the Becker daily water draws profile (Figure~\ref{fig:hot_water})~\cite{UofF_WaterDraws}. Water heating energy demands account for approximately \SI{14}{\percent} of total household electricity use in Switzerland~\cite{EnvironSwiss2013}. This corresponds to an average hot water demand of \SI{1.03}{\kilo\watt\hour} per household per day. The \acp{EWH} considered are rated at \SI{4.5}{\kilo\watt}. They are charged two to four hours each day and have a storage capacity of 24 hours. One \ac{EWH} is needed at each node to meet the hot water demands of the households\footnote{Multiple households may be in one building, and not all water heating is done using electricity.}. For simplification, it is assumed that all households follow the same hot water usage profile and use identical amounts of hot water each day. 

\subsection{Cost Structure}
SwissIX spot market prices are given for every hour of the measurement period. For simplicity, they are assumed to be known to the controller when dispatching the flexible loads. Taking bidding strategies into account is beyond the scope of this paper, but would most likely not significantly affect the dispatch.
All \ac{PV} plant owners are assumed to be paid a fixed feed-in tariff of \SI{0.36}{\EUR \per\kilo\watt\per\hour}.
Because the installed \ac{PV} capacity in Switzerland is too low to cause any \ac{PV} curtailment, the rules for \ac{PV} shedding during peak hours are based on the German Renewable Energy Resources Act~\cite{EEG2013}: the DSO is to compensate plant owners for \SI{95}{\percent} of their lost income plus any additional expenses incurred, minus any expenses saved, due to switching actions in the event of voltage instability. This penalty is necessary to provide an economic incentive for utilities to charge \acp{EWH} during peak price periods.

\begin{figure}
	\centering
	\includegraphics{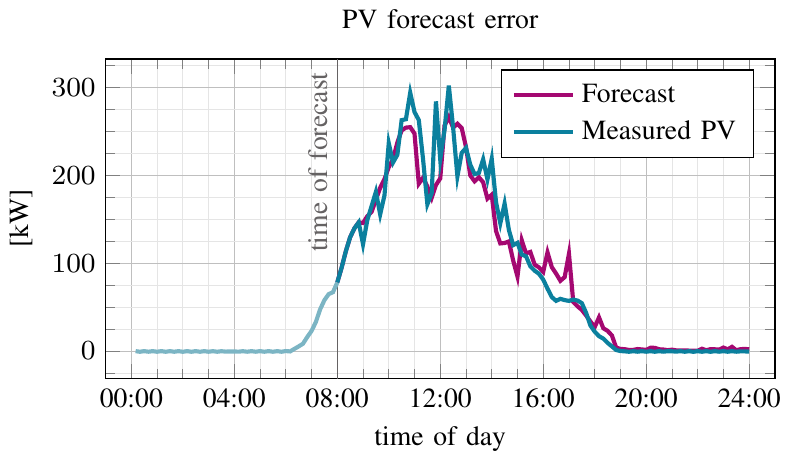}
	\label{fig:step4}
	\caption{\ac{PV} forecast at hour 8}\label{fig:prediction_error}
\end{figure}

\subsection{\ac{PV} Forecast}\label{sec:PV}
\ac{PV} prediction errors are introduced to study the effects of imperfect information. 
A random, normally distributed error factor is generated at every time step and multiplied with the smoothed measured data curve. The mean of the error factor increases linearly up to a maximum intraday mean of \SI{20}{\percent}~\cite{Lenzi2013PES}. The standard deviation is set to allow for a maximum overestimation at any given time of \SI{40}{\percent} and a maximum underestimation of \SI{30}{\percent}~\cite{Lorenz2010PVSEC}. Errors for all past time steps are set to zero, and a noise factor is added~\cite{Oldewurtel2010ACC}. 
Predictions can be updated hourly with increasing accuracy. 

\section{Increasing Hosting Capacity with DR}\label{sec:control}
\begin{figure}
  \centering
  \includegraphics{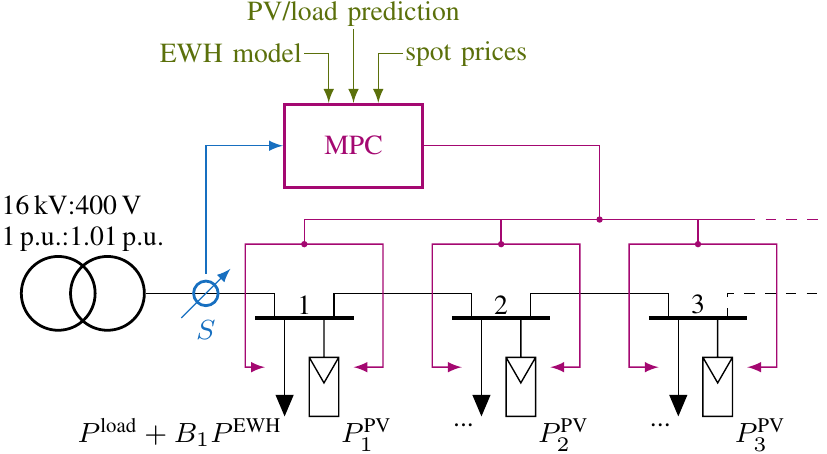}
  \caption{Proposed control scheme: Power is measured at the LV transformer (blue); the controller then solves an optimization problem based on forecasts (green) and sends control signals (purple) to flexible loads and PV installations.}
  \label{fig:scheme}
\end{figure}
We propose to only measure power backflows at the transformer. Limiting the backflows to a predefined level by directly switching flexible loads, and in extreme situations also some PV generation, should be sufficient in preventing over-voltages in the feeder as long as the distribution of loads and PV installations is known, and PV installations are not concentrated at a few nodes. This approach, schematically presented in Figure~\ref{fig:scheme}, minimizes the number of real-time measurements needed while providing sufficient dependability. 

Flexible loads and distributed generation must allow for \ac{DSO} control, but this controllability may be granular in power, i.e. only groups of assets can be switched. Delays in the control signal could be included in the framework, but is subject to future research. 

Flexible loads are optimally dispatched, taking into account electricity costs, asset constraints, and an estimate of the daily energy consumption of the \acp{EWH}. Not considering exact power flows and voltage levels in the controller simplifies calculations by eliminating non-linear power flow computations from the optimization. The linear program describing the load dispatch is explained in \ref{sec:opt}. Results of the dispatch are verified using historic load and PV data on the known grid topology (see Section~\ref{sec:description}), and possible voltage violations are identified. Finally, total costs are computed using spot market prices, actual load dispatch (including passive loads), and PV shedding penalties.

\subsection{Deterministic Optimization with Perfect Forecasts}\label{sec:opt}
The goal of the optimization is to minimize system costs within voltage and \ac{EWH} state of charge constraints. While it is cheaper to heat water during night hours, if excess PV production is expected, some or all of the water heating is shifted to daytime hours to avoid costly PV shedding penalties. The optimization is done with a temporal resolution, $\Delta t$, of 10 minutes as that is the relevant time scale for grid operators to take control action against voltage disturbances. The deterministic optimization problem formulation is as follows:
\begin{multline}
  \min_{B}\sum\limits_{k=1}^{N} \left(c^{\mathrm{spot}}_k \sum\limits_{i\in\mathcal{I}} P^\textrm{EWH}_i B_{i,k}\Delta t+c^{\mathrm{shed}} \sum_{s\in\mathcal{S}} P_{s,k}^\mathrm{shed}\Delta t \right.\\
    \left. + c^{\mathrm{sw}}\sum_{i\in\mathcal{I}}(S_{i,k}^{\mathrm{on}} + S_{i,k}^{\mathrm{off}})\right)\quad, 
  \label{eq:det_opt}
\end{multline}
s.t.\hfill$\forall k \in [1,N], \forall i \in \mathcal{I}, \forall s \in \mathcal{S}\qquad$

\begin{subequations}
  \begin{align}
    \begin{split}
      P^\mathrm{min} &\leq P^\mathrm{load}_k - P^\mathrm{PV}_k + \sum_{s\in\mathcal{S}} P_{s,k}^\mathrm{shed} \\
        &\qquad\qquad\qquad\qquad \sum_{i\in\mathcal{I}} P^\textrm{EWH}_i B_{i,k} \leq P^\mathrm{max}\quad,
    \end{split} \label{eq:1} \\    
    \SoC[i,k+1] &= \SoC[i,k] + P^\textrm{EWH}_i B_{i,k}\Delta t - \xi_k\quad, \label{eq:2}\\
    \SoC[i,1]&= \SoC[0]\quad, \label{eq:3}\\
    0 &\leq \SoC[i,k] \leq \SoC[\mathrm{max}]\quad,\label{eq:4} \\
    \SoC[i,N] &\geq \xi_1\quad,  \label{eq:5}\\
    0 &\leq P_{k}^\mathrm{shed} \leq P^{\mathrm{PV}}_{s,k}\quad, \label{eq:6} \\
    B_{i,k} &= B_{i,k-1} + S_{i,k}^{\mathrm{on}} - S_{i,k}^{\mathrm{off}}\quad. \label{eq:7}
  \end{align}
\end{subequations} 

The first term in~\eqref{eq:det_opt} minimizes the cost of water heating: $c^{\mathrm{spot}}$ corresponds to spot market prices, $ \mathcal{I}$ is the set of all \acp{EWH} or \ac{EWH} groups, $P^\textrm{EWH}_i$ is the power rating of \ac{EWH} $i$, and $B_{i,k}$ is the binary variable that specifies whether \ac{EWH} $i$ is on or off during time step $k$. The second term minimizes \ac{PV} power curtailment: $c^{\mathrm{shed}}$ represents the \ac{PV} shedding costs, $\mathcal{S}$ is the set of all PV installations or groups, and $P^\mathrm{shed}_s$ is the power shed by \ac{PV} unit $s$ during $\Delta t$. Finally, the last term penalizes each switching action $S^\mathrm{on}$ or $S^\mathrm{off}$ with a cost of $c^\mathrm{sw}$. $S_{i,k}^\mathrm{on}$ is one if \ac{EWH} $i$ is switched from off to on during time step $k$, and zero otherwise. $S_{i,k}^\mathrm{off}$ is defined analogously.

Constraint~\eqref{eq:1} sets power flow limits at the transformer: the power used by passive loads plus any electric water heating minus the net \ac{PV} power the system is able to accept must be within system limits. The positive limit is determined by the yearly peak-load in the double feeder (i.e. \SI{220}{\kilo\volt\ampere}), while the negative limit is set at a level that prevents over-voltages in ordinary simulation scenarios. 

Constraints~\eqref{eq:2}~-~\eqref{eq:5} control the energy behaviour of the \acp{EWH} in the system: The state of charge of each \ac{EWH} at time step k+1 is equal to the state at time step k plus any water heating done minus any water drawn~\eqref{eq:2}. The initial state of charge must be set~\eqref{eq:3}. The maximum capacity of each \ac{EWH} is limited~\eqref{eq:4}. The final state of charge at the end of the day must be large enough to supply the water needs for the first time step of the next day~\eqref{eq:5}.

Constraint~\eqref{eq:6} limits the amount of \ac{PV} curtailed by the actual level of \ac{PV} generation during the current time step, and the final constraint~\eqref{eq:7} affects switching behaviour: whether an \ac{EWH} is on or off at time step $k$ depends on its past state and whether or not there was any switching action. 

The \ac{EWH} dispatch $B_{i,k}$ is then applied to the system. Voltage evolution and losses in the feeder are simulated in minute resolution using the data described in Section~\ref{sec:description}. The deterministic optimization with perfect forecasts sets the benchmark for the lowest achievable costs and the minimum power curtailment levels in the system. Results form a basis for comparison with other scenarios. 

\subsection{Grouping of Control Signals}
Optimal system behaviour results when \acp{EWH} and \ac{PV} installations are controlled individually. However, this level of control is currently not implemented in Switzerland, where all \ac{EWH} loads of a distribution company may be controlled in only two or three groups. To investigate the effects of limited control on power curtailment and costs, $\mathcal{I}$ and $\mathcal{S}$ are varied while the $\SoC$, $w$, and $\xi$ variables are adjusted accordingly.
When \ac{PV} installations are individually controlled, power outputs of each installation is ramped down to the maximum allowed limit. When control is grouped, individual installations are turned off completely when over-voltages occur.  
Loads and \ac{PV} installations in one group are spread out geographically  so no section of the line is heavily affected when a group is switched in or out.

\subsection{Switching Penalty}
Increased switching of flexible loads in order to accommodate additional \ac{PV} capacity can have a negative impact on the lifetime of certain devices (e.g. heat pumps and air conditioners). The virtual cost $c^{sw}$ is varied to find the trade-off between energy loss and switching behaviour. 

\subsection{Model Predictive Control Formulation} 
A model predictive control algorithm is implemented to account for limited \ac{PV} forecast accuracy. The forecast uncertainty is created artificially as described in \ref{sec:PV}. Optimization is performed at every full hour, with a time horizon running until midnight, thus decreasing by one hour each time the optimization is run. The horizon resets at the beginning of the next day. The time horizon is not kept constant due to the generation patterns of \ac{PV}: there is no production at night and spot market prices are not know a day in advance.

The optimization problem is adjusted accordingly, and Equation~\eqref{eq:det_opt} becomes:
\begin{multline}
	\min_B \sum\limits_{k=0}^{N-1} \left( c_{t+k|t}^{\mathrm{spot}} \sum_{i\in\mathcal{I}} P^\textrm{EWH}_i B_{i,t+k|t}\Delta t + c^{\mathrm{shed}} \cdot\right.\\ \cdot  \sum_{s\in\mathcal{S}} P_{s,t+k|t}^\mathrm{shed}\Delta t
	\left. + c^\mathrm{sw}\sum_{i\in\mathcal{I}}\left(S_{i,t+k|t}^{\mathrm{on}} + S_{i,t+k|t}^{\mathrm{off}}\right)\right)~,
\end{multline}
with $N-1$ as the number of time steps until the end of the day, $t$ as the time of execution and with $x_{t+k|t}$ denoting the prediction of $x$ at time $t$, $k$ steps into the future. The first $B_{i,t+k|t}$ is applied; then new PV prediction is available and the optimization is solved once again.
Past time steps no longer need to be considered as long as the initial state of charge of the \acp{EWH} at time step $k$ is set to the final state of charge at the end of the previously implemented time step,
\begin{align}
  \SoC[i,t+0|t] &= \hat{x}^\textrm{SoC}_t & \forall i \in \mathcal{I}\qquad.
\end{align}

\section{Results}\label{sec:results}
%
%
%

The following definitions from~\cite{Bucher2013PVSEC} are used in this paper:

\begin{LaTeXdescription}
  \item[\ac{PV} Penetration] Total yearly solar energy fed into the grid divided by the yearly energy consumption of all consumers connected to the grid.
  \item[Hosting capacity] Maximum \ac{PV} penetration at which the \SI{3}{\percent} voltage rise limit is not violated. 
\end{LaTeXdescription}

\subsection{Deterministic Optimization Results}
Table~\ref{tab:penetration} shows the changes in the hosting capacity of the test grid under different \ac{PV} integration methods. DACHCZ regulation assumes zero system load with all \ac{PV} installations running at maximum power. Load correlations take into account actual \ac{PV} generation and consumption levels, while demand response introduces an additional flexible capacity of \SI{360}{\kilo\watt\hour} per day into the system. Hosting capacity is found assuming a uniform distribution of \ac{PV} in-feed at every node in the grid. Comparing \ac{DR} to the DACHCZ regulation gives an increase of nearly \SI{25}{\percent} points, effectively doubling the hosting capacity. However, it would be more fair to compare the \ac{DR} approach with load correlation, as both run under similar assumptions. Even here, over \SI{12.5}{\percent} points can be added: a sizable gain.

\begin{table}[htbp]
	\caption{\ac{PV} hosting capacities under different conditions}
	\centering
	\begin{tabular}{ccc} \toprule
		& \ac{PV} Hosting Capacity & Capacity per Household\\ \midrule
		DACHCZ & \SI{28.57}{\percent}  & \SI{0.865}{\kilo\watt_p}\\
		Load Correlation & \SI{39.97}{\percent}& \SI{1.209}{\kilo\watt_p}\\
		Demand Pesponse &\SI{52.78}{\percent} &\SI{1.597}{\kilo\watt_p}\\ \bottomrule
	\end{tabular}
	\label{tab:penetration}
\end{table}

\begin{figure}[htb]
        \centering
        \includegraphics{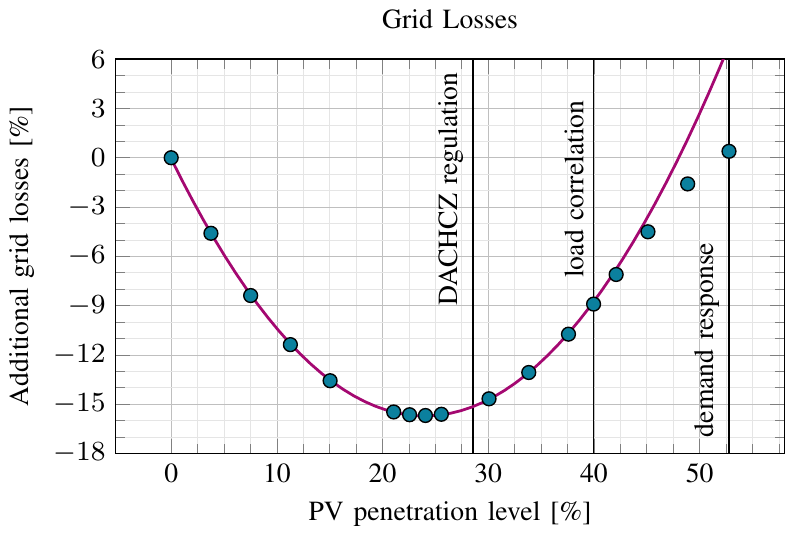}
        \caption{System losses at different \ac{PV} penetration levels}
        \label{fig:losses}
\end{figure}

Figure~\ref{fig:losses} shows how system losses behave at different \ac{PV} penetration levels. Losses here refer only to the losses on the \ac{LV} feeder. Losses in the transformer and at higher voltage levels are not regarded. At a \ac{PV} penetration level of \SI{0}{\percent}, grid losses account for \SI{1.81}{\percent} of the total power delivered during the year. Losses are minimized at a penetration level of \SI{24.1}{\percent}. The vertical lines give the hosting capacity with the three approaches. \ac{DR} is only used when penetration surpasses \SI{40}{\percent}. The purple line is a fit of all points below this level. While one would expect increasing losses for higher penetration levels due to the shifting of demand to peak production hours, losses can be kept at levels similar to the no-PV case. Furthermore, simulations not shown in the graph suggest that maximum losses will never significantly exceed initial system loss levels, as backflows are limited and PV shedding will occur before additional losses are generated in the LV feeder.


\subsection{Effects of Grouped Control Signals}
\begin{figure}[htb]
        \centering
        \includegraphics{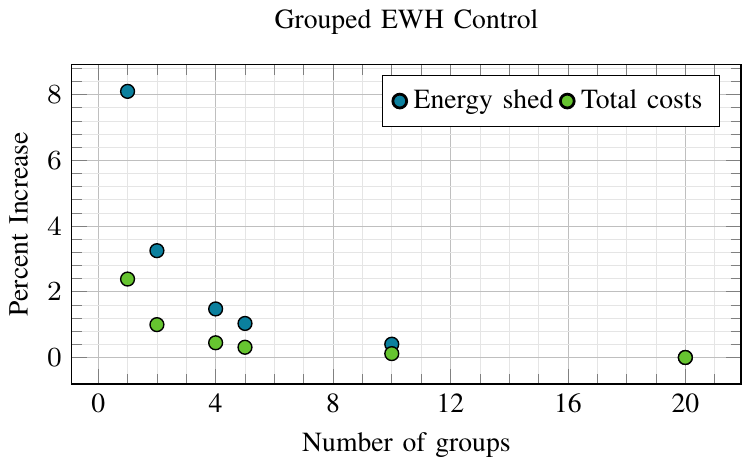}
        \caption{Increases in energy shed and total costs due to load grouping}
        \label{fig:load_grouping}
\end{figure}

\begin{figure}[htb]
        \centering
        \includegraphics{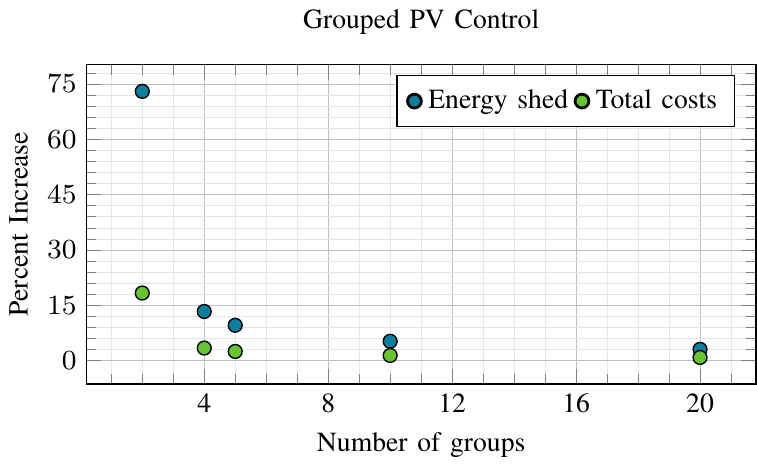}
        \caption{Increases in energy shed and total costs due to \ac{PV} grouping}
        \label{fig:pv_grouping}
\end{figure}

The effects of grouped control signals for \acp{EWH} and \ac{PV} installations are explored at a \ac{PV} penetration level of \SI{70}{\percent}, as considerable PV shedding occurs at this level. For lower levels of penetration, the effects of grouping would be less pronounced. Results are shown in Figures~\ref{fig:load_grouping} and~\ref{fig:pv_grouping}. In each case, the twenty \acp{EWH} and twenty \ac{PV} installations are evenly divided into the specified number of control groups. 

When \ac{EWH} control is grouped, energy shed and total system costs increase at almost identical rates. \ac{EWH} charging costs increase as more \acp{EWH} than necessary must be turned on during peak periods to accommodate the PV in-feed. This in turn causes state of charge levels to rise, resulting in a lower capacity to accommodate PV at other times of the day.

When the control of \ac{PV} installations is grouped and units can no longer be ramped, energy shed increases much more dramatically as whole installations must now be turned off even if current power levels are only marginally above system limits. Costs increase at a lower rate because as more PV is turned off, more \ac{EWH} charging can be shifted to lower cost periods. The decrease in charging costs helps to offset the increase in energy shedding costs. 

\subsection{Effects of Switching Penalties}
\begin{figure}[htb]
        \centering
        \includegraphics{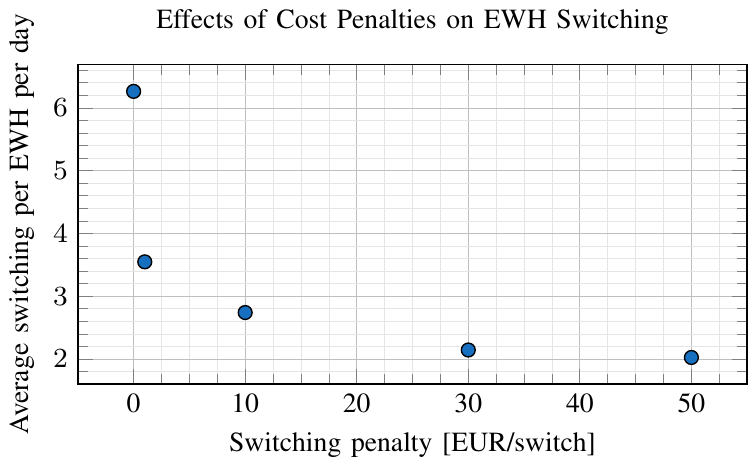}
        \caption{Reduced EWH switching through cost penalties}
        \label{fig:s_penalty1}
\end{figure}

\begin{figure}[htb]
        \centering
        \includegraphics{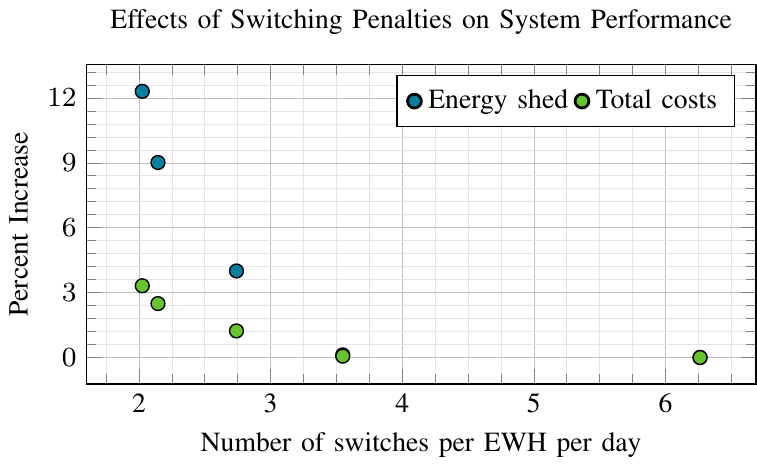}
        \caption{Increases in energy shed and total costs due to switching penalties}
        \label{fig:s_penalty2}
\end{figure}

The switching penalty is an arbitrary value set in relation to the PV shedding costs, assumed here to be \SI{34.2}{ct}/kWh. Intuitively, as the switching penalty increases, the amount of switching decreases (Figure~\ref{fig:s_penalty1}) while energy shedding and total system costs increase (Figure~\ref{fig:s_penalty2}). 

For comparison purposes, each \ac{EWH} is switched an average of 4.47 times a day when the system is optimized to heat water at the lowest spot prices, with no consideration for PV shedding. The nominal switching value per \ac{EWH} per day is two: once to turn on and once to turn off. A small switching penalty has a large initial effect, as it removes any unnecessary switching the \acp{EWH} do between periods of similar spot prices. Higher penalties must consider the trade-off between energy shed / cost increases and the wear and tear on the equipment due to switching.

\subsection{Effects of Forecast Inaccuracies} 
The effects of PV forecast inaccuracies are shown in Table~\ref{tab:mpc}. Using day-ahead forecasts, the \ac{EWH} charging schedule for the upcoming 24-hour period is set each day at 00:00. Then hourly forecasts are assumed to be available, and the model predictive control algorithm updates the charging schedule of the EWHs at the start of each hour based on more accurate predictions. The performance of these two methods are compared against the deterministic optimization reference case. When faced with uncertainties, the amount of energy that is shed is highly increased. As expected, an hourly update can considerably reduce these losses. While it is conceivable to send a daily switching plan to each \ac{EWH}, it seems much more promising to have real-time switching capabilities.

\begin{table}[htbp]

	\caption{Effects of \ac{PV} Forecast Inaccuracies}
	\centering
	\begin{tabular}{ccc} \toprule
		& Increase in Energy Shed & Increase in Total Costs \\ \midrule
		Hourly Forecast Updates & \SI{4.17}{\percent}  & \SI{1.07}{\percent}\\
		Day-ahead Forecast & \SI{29.61}{\percent}& \SI{7.61}{\percent}\\
 \bottomrule
	\end{tabular}
	\label{tab:mpc}

\end{table}

\section{Discussion}\label{sec:discussion}
This paper proposed and investigated a simple control architecture that increases the \ac{PV} hosting capacity of a \ac{LV}-feeder with \ac{DR}. The controller only relies on real-time power flow measurements at the \ac{LV} transformer, thus minimizing investment costs in measuring infrastructure. Results are promising, as the hosting capacity of the test system can be substantially increased from \SI{28.57}{\percent} with current regulation to over \SI{50}{\percent} when flexible loads are used to absorb power locally. As opposed to methods relying on reactive power management for voltage compliance, no unnecessary currents are injected into the feeder. Furthermore, losses can be slightly reduced due to better correlation between local production and consumption.

Grouping of both flexible loads and PV installations was investigated. While control performance is reduced as the number of groups is reduced, and thus increasing the granularity of the control action, it was shown that full or continuous controllability of assets is unnecessary. This is a relevant result, as currently installed communication channels for load switching, such as ripple-control in Switzerland, only allow for the switching of load groups, and many existing PV converters can only be turned on or off even though they should technically be able to continuously set power output.

Minimal shedding of PV power is achieved when loads can be switched at any time, but this may be undesirable from a life-time perspective and may alienate customers from participating in such a \ac{DR} scheme. However, most of the switching adds little benefit to overall system performance. By adding a virtual penalty, switching can be reduced to an acceptable level of less than four times a day, i.e., activating a load less than twice a day.

Despite these encouraging findings, there are challenges in the implementation of the approach. Firstly, backflow limits need to be identified. This can be done heuristically by running simulations of typical years. For this, PV in-feed and demand profiles of local customers, as well as the locations of the PV installations must be known or estimated. One may also use analytic approaches, considering installed peak power and minimum expected load. Note, we assumed evenly distributed loads and PV installations. If the PV production is concentrated at one node, there may be current violations within the feeder even if power consumption and production is balanced. This cannot be handled by the presented approach.

To optimally use the available flexibility, the daily energy demand of the flexible loads must also be known. For this, dedicated measurements of \acp{EWH} at customer premises must be available. These measurements do not need to be communicated in real time and only need to be given with a time resolution in days, but even this is currently unavailable. It would be straight forward to gather this data with an \ac{AMI}, but the \acp{SM} currently rolled out often do not have dedicated measuring capabilities for switchable loads, requiring a second meter to be installed.

Finally, PV forecasts were subject to severe uncertainty. \ac{MPC} approaches can handle a good part of the uncertainty, but still leave room for improvement. Techniques from robust control and stochastic control should be investigated, and it is the authors' belief that improvements are still achievable. Robust control would shift most flexible loads to the day, unnecessarily increasing costs but guaranteeing minimal shedding. Stochastic control, either using probability constraints or a scenario-based approach to minimize expected costs, might be the favorable solution: the problem as such does not require the strict guarantees offered by robust control -- PV shedding is an undesirable but always feasible strategy.

\section{Conclusion and Outlook}\label{sec:conclusion}
This paper explores the technical potential of \ac{DR} for integration of \ac{PV} generation in \ac{LV} grids and shows its capabilities using a simple control method: only power flow at the transformer is measured in real-time; all other relevant data, such as the daily energy consumption of \acp{EWH}, can be gathered using an \ac{AMI}. Granular control of assets was shown to be sufficient, meaning a legacy control infrastructure, such as ripple-control can be used. The hosting capacity of the feeder was increased to over \SI{50}{\percent} while loads were activated less than twice a day. 
Further studies should look at stochastic control to consider the inherent uncertainty in the optimization inputs. A more accurate model for hot water consumption, ideally based on real measurements, should be implemented to account for seasonal demand variations.
In this context, it would be interesting to see comparable analyses for regions with higher PV capacity factors and that use air-conditioning as flexible loads (e.g. California, southern Europe or South-East Asia).
Furthermore, to gauge whether \ac{DR} can make a feasible business case, the study must be explored from a regulatory point of view and cost incentives must be clearly identified. 

\section{Acknowledgment}
The authors gratefully received grid data and PV production time-series from project VeIN, and household load data gathered by ewz. Theodor Borsche is financed by the project \emph{Distributed Demand Response}, which is funded by the CTI (Commission for Technology and Innovation).

\bibliographystyle{IEEEtran}
\bibliography{PV_integration_with_DR}

\begin{thebibliography}{10}
\providecommand{\url}[1]{#1}
\csname url@samestyle\endcsname
\providecommand{\newblock}{\relax}
\providecommand{\bibinfo}[2]{#2}
\providecommand{\BIBentrySTDinterwordspacing}{\spaceskip=0pt\relax}
\providecommand{\BIBentryALTinterwordstretchfactor}{4}
\providecommand{\BIBentryALTinterwordspacing}{\spaceskip=\fontdimen2\font plus
\BIBentryALTinterwordstretchfactor\fontdimen3\font minus
  \fontdimen4\font\relax}
\providecommand{\BIBforeignlanguage}[2]{{%
\expandafter\ifx\csname l@#1\endcsname\relax
\typeout{** WARNING: IEEEtran.bst: No hyphenation pattern has been}%
\typeout{** loaded for the language `#1'. Using the pattern for}%
\typeout{** the default language instead.}%
\else
\language=\csname l@#1\endcsname
\fi
#2}}
\providecommand{\BIBdecl}{\relax}
\BIBdecl

\bibitem{EPIA2013}
G.~Masson, M.~Latour, M.~Rekinger, L.-T. Theologitis, and M.~Papoutsi,
  ``{Global Market Outlook for Photovoltaics 2013-2017},'' European
  Photovoltaic Industry Association, Tech. Rep., 2013.

\bibitem{REN21_2013}
{REN 21 Steering Committee}, ``{Renewables 2013 Global Status Report},''
  Renewable Energy Policy Network for the 21st Century, Paris, Tech. Rep.,
  2013.

\bibitem{Bloomberg2013}
\BIBentryALTinterwordspacing
J.~Isola and A.~Mccrone, ``{Solar to Add More Megawatts than Wind in 2013 , for
  First Time},'' New York City, Sep. 2013. [Online]. Available:
  \url{https://www.bnef.com/PressReleases/text/324}
\BIBentrySTDinterwordspacing

\bibitem{Ernst2012}
B.~Ernst and B.~Engel, ``{Grid Integration of Distributed PV-Generation},'' in
  \emph{IEEE PES General Meeting 2012}.\hskip 1em plus 0.5em minus 0.4em\relax
  San Diego: IEEE, 2012.

\bibitem{Degner2011}
T.~Degner, F.~I. Germany, S.~M.~A. Solar, T.~Ag, G.~Arnold, M.~Breede,
  T.~Reimann, and S.~A, ``{Increasing the Photovoltaic-system Hosting Capacity
  of Low Voltage Distribution Networks},'' in \emph{21st International
  Conference on Electricity Distribution}, no. 1243, Frankfurt, 2011, pp. 6--9.

\bibitem{Bucher2013PVSEC}
C.~Bucher, G.~Andersson, and L.~K\"{u}ng, ``{Increasing the PV Hosting Capacity
  of Distribution Power Grids -- A Comparison of Seven Methods},'' in
  \emph{28th European Photovoltaic Solar Energy Conference and Exhibition
  (PVSEC)}, Paris, 2013, pp. 4231 ---- 4235.

\bibitem{DACHCZ2007}
J.~M. {Gerhard Bartak, Hansj\"{o}rg Holenstein}, ``{Technical Rules for the
  Assessment of Network Disturbances, 2nd ed., Austria, Germany, Czech
  Republic, Switzerland},'' Tech. Rep., 2007.

\bibitem{UofF_WaterDraws}
P.~Fairey and D.~Parker, ``{A Review of Hot Water Draw Profiles Used in
  Performance Analysis of Residential Domestic Hot Water Systems},'' Florida
  Solar Energy Center / University of Central Florida, Cocoa, Tech. Rep., 2004.

\bibitem{EnvironSwiss2013}
{Environment Switzerland}, ``{State of the Environment - Households and
  Consumption},'' Tech. Rep., 2013.

\bibitem{EEG2013}
EEG, ``{Act on Granting Priority to Renewable Energy Sources},'' Tech. Rep.,
  2008.

\bibitem{Lenzi2013PES}
V.~Lenzi, A.~Ulbig, and G.~Andersson, ``{Impacts of Forecast Accuracy on Grid
  Integration of Renewable Energy Sources},'' in \emph{IEEE PES PowerTech
  2013}.\hskip 1em plus 0.5em minus 0.4em\relax Grenoble: IEEE, Jun. 2013.

\bibitem{Lorenz2010PVSEC}
E.~Lorenz, T.~Scheidsteger, J.~Hurka, D.~Heinemann, and C.~Kurz, ``{Regional PV
  Power Prediction for Improved Grid Integration},'' in \emph{25th European
  Photovoltaic Solar Energy Conference and Exhibition (PVSEC)}, vol.~19, no.
  Appl. 2011, Valencia, 2010, pp. 757--771.

\bibitem{Oldewurtel2010ACC}
F.~Oldewurtel, A.~Parisio, C.~N. Jones, M.~Morari, D.~Gyalistras, M.~Gwerder,
  V.~Stauch, B.~Lehmann, and K.~Wirth, ``{Energy Efficient Building Climate
  Control Using Stochastic Model Predictive Control and Weather Predictions},''
  in \emph{American Control Conference}, Baltimore, 2010, pp. 5100--5105.

\end{thebibliography}

\end{document}